\documentclass{JHEP3}
\usepackage{epsfig,cite}
\usepackage{amsmath,amstext}

\title{Saxion Emission from SN1987A}
\author{Daniel Arndt\thanks{arndt@phys.washington.edu}, 
        Patrick J. Fox\thanks{pjfox@phys.washington.edu}\\
        Department of Physics, Box 1560, University of Washington,\\
        Seattle, WA 98195-1560, USA
       }
\abstract{
We study the possibility of emission of the saxion, 
a superpartner of the axion,
from SN1987A.  
The fact that the observed neutrino
pulse from SN1987A is in excellent agreement with the current theory
of supernovae places a strong bound on the energy loss into
any non-standard model channel, 
therefore enabling bounds to be placed on the decay
constant, $f_a$, of a light saxion.
The low-energy coupling of the saxion, 
which couples at high energies to the QCD gauge field strength, 
is expected
to be enhanced from QCD scaling,
making it interesting to investigate if the saxion could place
stronger bounds on $f_a$ than the axion itself.
Moreover, since the properties of the saxion
are determined by $f_a$, a constraint on this parameter
can be translated into a constraint on the 
supersymmetry breaking scale.
We find that the bound on $f_a$ from saxion emission
is comparable with the one derived from axion emission due to a
cancellation of leading-order terms in the soft-radiation expansion.
}

\preprint{UW-PT-02-11, NT@UW-02-0014}

\begin{document}

\unitlength=1mm

\def\hsp  {\hspace{ 0.27em}}
\def\nhsp {\hspace{-0.27em}}
\def\hspp {\hspace{ 0.18em}}
\def\nhspp{\hspace{-0.18em}}
\newcommand{\nott}[1]{{\hspp\not{\nhsp #1\hsp}\nhspp}}
\newcommand{\Nbar}{\ensuremath{\bar{N}}}
\newcommand{\gfive}{\ensuremath{\gamma_5}}
\newcommand{\be}{\begin{equation}}
\newcommand{\ee}{\end{equation}}
\newcommand{\ber}{\begin{eqnarray}}
\newcommand{\eer}{\end{eqnarray}}
\newcommand{\mbf}{\mathbf}
\newcommand{\pplus}{\ensuremath{p_{+}}}
\newcommand{\pminus}{\ensuremath{p_{-}}}
\newcommand{\pprime}{\ensuremath{p'}}
\newcommand{\pprimeprime}{\ensuremath{p''}}
\newcommand{\pqsym}{{U(1)_\text{PQ}}}
\newcommand{\fa}{{f_{a}}}
\newcommand{\Dsl}{{D\!\!\!\!\slash}}
\newcommand{\dsl}{{\partial\!\!\!\slash}}
\def\a{{\alpha}}
\def\b{{\beta}}
\def\d{{\delta}}
\def\e{{\epsilon}}
\def\g{{\gamma}}
\def\k{{\kappa}}
\def\l{{\lambda}}
\def\m{{\mu}}
\def\n{{\nu}}
\def\thCM{{\theta_{\mbox{cm}}}}
\def\o{{\omega}}
\def\O{{\Omega}}
\def\th{{\theta}}
\def\sl{{\!\!\!\slash}}
\def\Tr{{\text{Tr}}}

\section{Introduction}\label{sec:intro}
Searches for the neutron electric dipole moment (EDM) $d_n$
place strong constraints on
the QCD vacuum angle  $\theta_\text{QCD}$%
~\cite{Harris:1999jx}.
The latest
incarnation of these experiments gives an upper bound of
$|d_n|<6.3\times10^{-26}e$~cm
which can be related~\cite{Crewther:1979pi} 
to the CP-violating parameter in QCD, $\theta_{QCD}$,
to give 
$\theta_{QCD}<(1-10)\times 10^{-10}$.
  Bounds on $\theta_{QCD}$ coming from $^{199}\text{Hg}$ EDM
  experiments~\cite{Romalis:2000mg} are somewhat stronger, 
  $\theta_{QCD}<1.5\times 10^{-10}$.  
  However,
  the conversion of the bound on $d(^{199}\text{Hg})$ 
  to a constraint on $\theta_{QCD}$ is
  model-dependent and subject to large uncertainties.

A possible explanation%
~\cite{Peccei:1977hh,Peccei:1977ur,Dine:1981rt,Kim:1979if,Shifman:1980if} 
for the smallness of $\theta_\text{QCD}$
is the introduction of a new anomalous global $\pqsym$ symmetry
which is spontaneously broken at the Peccei-Quinn (PQ) scale $\fa$.  
Associated with this spontaneously broken symmetry is
a new particle, called the
axion~\cite{Weinberg:1978ma,Wilczek:1978pj}. 
The axion is the pseudo Goldstone boson of the broken PQ symmetry.  It
develops a mass 
$m_a\simeq m_\pi f_\pi/f_a$
due to the
axial anomaly.

The present constraints on $f_a$ are $10^9~\text{GeV} \lesssim \fa \lesssim
10^{12}~\text{GeV}$.
The
lower bound on the allowed window comes from SN1987A and corresponds to an
axion whose coupling is such that it can be abundantly produced in
the nascent neutron star and yet interacts weakly enough to allow emission on
a similar time scale to the neutrinos.  This axion provides a new efficient
cooling mechanism and would lead to
a shortening of the neutrino pulse from SN1987A, in conflict with
observations, if $f_a$ is too small~\cite{Raffelt:1996wa}.   
The upper limit on the allowed window
comes from cosmological arguments. It corresponds to axions so weakly coupled
that they never thermalized in the early universe.  
At high temperatures the axion
field value is misaligned  with the value it takes at low temperatures and as
the universe cools the axion begins to ``roll down" towards its
minimum~\cite{Kolb:1990vq}.  This leads to coherent oscillations of the axion
field.  The requirement that the energy density in these oscillations does not
overclose the universe leads to the upper bound on $\fa$.

In this paper we are interested in models which are  supersymmetric and use
$\pqsym$ to solve the strong CP problem.   If supersymmetry (SUSY) is combined
with $\pqsym$ there are superpartners to the axion, which are the fermionic
axino and the bosonic saxion.   For a given axion model the saxion's
properties are determined by SUSY.
In particular, for gauge mediated models%
~\cite{Asaka:1998ns,Asaka:1998xa}, the saxion mass $m_s$
is comparable to the gravitino mass 
$m_{3/2}\sim10^{-2}~\text{eV}-1~\text{GeV}$.
The saxion can be light and could be 
emitted from the interior of a
supernova where the temperature is of order 20~MeV.
  
We find that the saxion couples in a fashion similar to the
dilaton~\cite{Kaplan:2000hh}:
It couples to the QCD gauge field
strength in the ultraviolet (UV) and this coupling
is enhanced in the infrared (IR) due to QCD scaling.
This enhancement leads to the possibility that astrophysical 
bounds on $f_a$ from the saxion could be
stronger than those from the axion.
In certain gauge mediated models of SUSY breaking
(see \cite{Asaka:1998xa}, for example)
the PQ scale, $f_a$, is derived from the SUSY
breaking parameters.
Any constraints on $f_a$ can be turned into constraints
on these SUSY breaking parameters.

Supernovae (SN), in particular the recent SN1987A, have been used frequently
to obtain bounds on new physics using the energy-loss
argument.
Our current theory of SN explains the shape and duration of SN1987A's
neutrino pulse and is in agreement with the observed data.  
If there exists a new channel 
that competes with the neutrinos and transports a comparable
amount of energy from the interior of the SN
then the current description of SN1987A's 
neutrino signal is significantly altered.
A criterion for the 
maximum possible emissivity of a 
novel 
energy loss process
has been given by Raffelt~\cite{Raffelt:1996wa}.
It states that the
luminosity of the new particle(s) can not
exceed the neutrino luminosity of 
$L_\nu\sim 10^{52}$~ergs/s.  
For a proto-neutron star as developed during SN1987A this criterion
translates to an 
emissivity bound of  
$\dot{\epsilon}\le 10^{19}$~ergs/g/s
for the new channel, the saxion in our case. 
Saxion parameters that lead to an emissivity greater than this are ruled out.

Several calculations of axion
emissivity from the process $NN\to NNa$ have 
constrained $\fa$~%
\cite{Iwamoto:1984ir,Brinkmann:1988vi,Burrows:1989ah,
      Hanhart:2000ae,Iwamoto:1992jp}.
Until recently, the NN scattering amplitude
was derived from a one-pion-exchange (OPE) potential.
However, as shown in \cite{Hanhart:2000ae}, it is possible
to relate the amplitude for axion emission
to the on-shell NN scattering amplitude which can be
derived from the experimentally measured NN scattering data.
In contrast to the OPE calculation, this method gives
a 
model-independent result. 

Naively one might expect that such a model independent
calculation could be repeated for the saxion emission channel $NN\to NNs$.  
This, however, is not the case due
to a cancellation of the leading order terms in the expansion around the soft
radiation limit. It is exactly these terms that make a model-independent
analysis possible.  

In the non-relativistic limit the saxion---being a scalar particle---%
couples to the nucleon mass.  Hence there is no dipole radiation.%
\footnote{This is similar to photon radiation from the 
collision of two charged particles with the same mass.
In this case the center of mass coincides with the center of
charge, there is no dipole radiation and no emitted bremsstrahlung.}
Axions, on the other hand, couple to spin, which is
not conserved in nucleon interactions and therefore they can be emitted 
by bremsstrahlung.  
Diagrammatically, the leading order poles
corresponding to emission from an external leg sum in the case of axions and
cancel in the case of saxions.  

For this reason we will calculate the amplitude
for saxion emission at tree level in the $SU(3)$ chiral Lagrangian.  Using
this Lagrangian to model the nucleon-nucleon interaction has several
shortcomings, as explained below, and one might doubt the validity of our
procedure.  However, given the rather large uncertainties in the astrophysical
parameters entering the calculation, it gives a sufficient estimate for the
emissivity.  

We find that
saxion emission is comparable to axion emission
suggesting that the bound on $f_a$ from saxions
could be at least
as significant as that from axions.%
\footnote{
  Here and throughout the paper we assume a light saxion, i.e.
  the mass of the saxion
  is
  less than the typical temperature in the neutron star
  ($T\simeq10-50$~MeV) and can be neglected.
}
However,
several simplifying approximations
were made
in order to carry out the calculation.
Once a model-independent calculation becomes feasible it would be
interesting to repeat this calculation without relying on 
these approximations.
Including the effects of this previously ignored emission channel
could tighten the constraints on the PQ parameter space,
if the UV physics is supersymmetric.

This paper is laid out as follows.  In Section~\ref{sec:saxions} we derive the
properties of the saxion at energies above the SUSY breaking scale.
In Section~\ref{sec:lowenergy} we compute the low
energy effective coupling of the saxion.
In Section~\ref{sec:emissivity} we 
discuss the soft radiation limit,
calculate the saxion emissivity
and derive bounds on $\fa$.
We conclude in Section~\ref{sec:conclusions}.

\section{(S)axions} \label{sec:saxions}
The gauge field piece of the standard model Lagrangian for a theory 
with a non-zero 
$\theta_\text{QCD}$-angle is%
\footnote{In the usual fashion, 
  $\tilde{F}_{\mu\nu}
  =
  \epsilon_{\mu\nu\alpha\beta}F^{\alpha\beta}/2$ 
and the trace is over group indices.}
\begin{equation} \label{eqn:lagrangian}
  {\mathcal L}
  =
  -\frac{1}{2g^2}\Tr\,F_{\mu\nu}F^{\mu\nu}
  -\frac{\theta_\text{QCD}}{16\pi^2}\Tr\,F_{\mu\nu}\tilde{F}^{\mu\nu}
,\end{equation}
where
\begin{equation} \label{eqn:thetaQCD}
  \theta_{\text{QCD}}= \theta+\text{Arg Det} M 
\end{equation}
and $g$ is the $SU(3)$ gauge coupling constant.
Classically,  
the bare $\theta$ angle can be removed from the QCD Lagrangian 
by a chiral
transformation of the quark fields.  
However, in the absence of a massless quark,
this transformation is no longer a symmetry of 
the quantum theory since chiral
transformations are anomalous. 
Therefore, 
$\theta$ gets shifted by
chiral phase rotations of the quark fields
whereas $\theta_{QCD}$ as defined in 
Eq.~(\ref{eqn:thetaQCD})
is invariant.

Consistency with experiment requires $\theta_\text{QCD}<10^{-9}$.
Explaining the smallness of $\theta_\text{QCD}$
is the strong $CP$ problem.
The PQ solution to the strong $CP$ problem
postulates the existence of a new dynamical field, the axion $a$,
which allows $\theta_\text{QCD}$ to vanish dynamically,
$\theta_\text{QCD}\rightarrow\theta_\text{QCD} + a(x_\mu)/\fa$.  
This is
achieved by introducing an anomalous global symmetry $U(1)_{PQ}$
spontaneously broken at the scale $f_a$.  

Realizing $\pqsym$ in a supersymmetric theory necessitates 
putting the axion in
a chiral supermultiplet.  
This multiplet is filled out 
by introducing its fermionic partner, the axino,
and making the axion $a$ 
the real part of a complex scalar field. 
The saxion $s$ 
is the imaginary part of this scalar field and
its couplings are uniquely determined by SUSY from those of the axion.

In a supersymmetric theory the Yang-Mills action is 
most conveniently written 
in superspace notation as
\begin{equation} \label{eqn:YMaction}
  {S}
  =
  \frac{1}{8\pi}
  \text{Im}\,
  \tau
  \int d^4x\,d^2\theta\,\Tr\,W^{\alpha}W_{\alpha}
,\end{equation}
where $\theta$ is a two-component Grassmann variable
and $W^{\alpha}$ is a vector superfield which
packages the gauge vector field $A_\mu$, its superpartner
the gaugino, and an auxiliary field.
The coupling parameter $\tau$ is chosen as
$\tau=\frac{4\pi i}{g^2} +\frac{\theta_\text{QCD}}{2\pi}$
so that upon expanding Eq.~(\ref{eqn:YMaction})
one recovers the terms in Eq.~(\ref{eqn:lagrangian}). 
Making $\theta_\text{QCD}$ dynamical
corresponds to 
$\tau\rightarrow\tau+\frac{\Phi}{2\pi\fa}$
where $\Phi$ is a chiral superfield containing
the axion, saxion, and
axino.
Expanding $\Phi$ in terms of its component fields gives a shift
of the bosonic piece:
$\tau\rightarrow\tau+\frac{1}{2\pi\fa}(a+is)$.  
The saxion $s$ couples to $F^2$
and the axion $a$ couples to $F\tilde{F}$, as before.

The axion is the pseudo Goldstone boson of $\pqsym$; it would be
massless were it not for the axial anomaly. 
Without SUSY the axion gets a vev below the PQ scale
and develops a `mexican hat' potential.
The radial mode becomes heavy 
(on the order of the symmetry breaking scale) and the
angular mode remains light, becoming the axion.
With unbroken SUSY the superpotential is holomorphic and this radial direction
remains flat, ignoring the effects of the anomaly.  There is now an extra light mode corresponding to oscillations
in the radial direction: the saxion.  
With unbroken SUSY there are two degenerate light modes
whose mass is set by the effects of the axial anomaly.  
Once SUSY is broken
this nearly flat direction 
is lifted and the saxion acquires a mass, independent of
the axion.  The specifics
of the SUSY breaking mechanism determine the 
form of the potential for $\Phi$ and thus
the masses of the saxion and axion.
Gauge mediated models predict viable properties for
the saxion: Its mass can be light, of 
${\mathcal O}(10^{-2}\text{eV}-1\text{GeV})$, 
and $f_a$ fits naturally in the
allowed window~\cite{Asaka:1998xa,Asaka:1998ns}.

\section{Low Energy Saxion Dynamics}\label{sec:lowenergy}
As we run down in energies 
from above the PQ scale, $f_a$, 
passing the SUSY breaking scale, $M_\text{SUSY}$,
to below the QCD scale, $\Lambda_\text{QCD}$,
we pass through various effective theories that
describe the relevant degrees of freedom at that energy.
Summarizing what was mentioned in the last 
section ($E\sim\sqrt{q^2}$, where $q$ is a typical momentum transfer):
\begin{itemize}
  \item
    $E>\fa,\ M_\text{SUSY}$:
    $\pqsym$ unbroken, SUSY unbroken.
  \item
    $f_a>E>M_\text{SUSY}$:
    $\pqsym$ broken, SUSY unbroken, degenerate saxion and
    axion with mass generated by the axial anomaly.
  \item
    $M_\text{SUSY}>E>\Lambda_\text{QCD}$:
    SUSY broken, superpartners
    become heavy, nondegenerate massive saxion and axion,  
    quarks and gauge fields are relevant degrees of freedom.
  \item
    $\Lambda_\text{QCD}>E$:
    Mesons, baryons, saxion, and axion are relevant degrees
    of freedom.
\end{itemize}

The Lagrangian for the saxion-axion system at $\fa$ is,
\begin{eqnarray} \label{eqn:bigL}
  {\mathcal L}
  &=& 
  \frac{1}{2}(\partial_{\mu}s)^2+\frac{1}{2}(\partial_{\mu}a)^2
  -V(a,s)
  -\left(\frac{1}{2g^2}+\frac{1}{16\pi^2}\frac{s}{\fa}\right)\Tr F^2
                \nonumber     \\
  &&
  -\frac{1}{16\pi^2}
   \left(\theta_\text{QCD}+\frac{a}{\fa}\right)
   \Tr F\tilde{F}
  +
  \sum_{i=1}^6
    \left(
      \bar{q}_i i\nott{D} q_i-m_i \bar{q}_iq_i
      +\frac{1}{\fa}\partial_{\mu}a\,\bar{q}_i\gamma^\mu\gamma_5 X q_i
    \right)
,\end{eqnarray}
where the quarks $q_i$ have mass $m_i$ with the index $i$
running over quark flavors.
Furthermore,
$V(a,s)$ is undetermined 
and
$X$ is a model
dependent flavor mixing matrix.

In order to compute the couplings of the saxion to matter 
at the low energies typical
for neutron stars we need to
first integrate out the heavy quarks.
We can then match this theory
onto a low-energy Lagrangian containing hadrons as the
relevant degrees of freedom.  
The saxion's interactions can be written in
terms of the renormalization group invariant quantities 
$m_i\bar{q}_i q_i$ and 
$\partial_{\mu}D^{\mu}$, 
where the latter is the divergence of the scale current.
This method of evaluating the divergence of the scale current in order
to calculate matrix elements of the $F^2$ operator
at low momentum transfer has been developed 
in~\cite{Shifman:1978zn,Voloshin:1987hp,Chivukula:1989ze,Chivukula:1989ds} 
and recently applied to systems involving 
the low energy coupling of the
dilaton~\cite{Kaplan:2000hh}
and quarkonium~\cite{Chen:1998zz}.

At the scale $\fa$, 
where the Lagrangian in Eq.~(\ref{eqn:bigL}) is relevant,
$\partial_\mu D^\mu$ is given by
\begin{equation}\label{eqn:dDbigL}
  \partial_\mu D^\mu
  =
  T^\mu_\mu
  =
  \frac{\beta}{g^3}\Tr F^2
  +\sum_{i=1}^6(1-\gamma_{m_i})m_i\bar{q}_i q_i
.\end{equation}
Here we have ignored the contributions from the 
$s F^2$ and $a F\tilde{F}$ terms
since they are suppressed by a factor of 
$1/\fa$ compared to the terms in 
Eq.~(\ref{eqn:dDbigL}).  
Furthermore, $\beta=\partial
g/\partial \log\mu$ is the $SU(3)$ beta function for the
coupling $g$ 
given to leading order by $\beta=-b_0g^3/16\pi^2$.
Above the SUSY breaking scale $b_0$ is given by 
$b_0=9-N_f$ for the Minimal Supersymmetric Standard Model.
Below the SUSY breaking scale it is given by the
standard model result
$b_0=11-2N_f/3$,
with $N_f$ being the number of flavors. 
Furthermore,
$\gamma_{m_i}=\partial m_i/\partial\log\mu$
is the mass anomalous dimension.
Using
Eq.~(\ref{eqn:dDbigL}) we can rewrite the saxion-gauge field piece of
Eq.~(\ref{eqn:bigL}) as,
\begin{equation} \label{eqn:Ls}
  {\mathcal{L}}_s
  =
  \frac{s}{16\pi^2\fa}
  \frac{g}{\beta}
  \left(
    \partial_\mu D^\mu
    -\sum_{i=1}^{6}(1-\gamma_{m_i})m_i\bar{q}_i q_i
  \right)
.\end{equation}

We can now run down in scale from $\fa$ 
to $\Lambda_\text{QCD}$ via $M_\text{SUSY}$, matching
across the quark mass thresholds as we go.  
Working to leading order in 
$\alpha_s=g^2/4\pi$ 
and neglecting $\gamma_m$, which is small compared to unity,  
we arrive at
\begin{equation}\label{eqn:L-s}
  {\mathcal L}_s
  =
  \frac{s}{16\pi^2\fa} 
  \kappa
   \left(\partial_\mu D^\mu-\sum_{i=1}^3 m_i\bar{q}_i q_i\right),
\end{equation}
where the enhancement factor $\kappa$ is given by
\begin{equation}\label{eqn:kappa}
  \kappa
  =
  \frac{4\pi/9}{\a_s(f_a)}
  \approx
  \frac{2}{3}
  \log\frac{f_a}{\text{GeV}}
  +12.4
.\end{equation}
This running gives sizable enhancements to the saxion
coupling.
For instance, 
for $f_a=10^9$~GeV, 
a SUSY breaking scale of $10^4$~GeV, and 
$\Lambda_\text{QCD}=200$~MeV
this QCD scaling
results in an enhancement of $\kappa\approx 26$ over the naive coupling.

\section{Calculating the Emissivity} \label{sec:emissivity}
As detailed in the introduction, the Raffelt criterion limits 
any mechanism that competes with the neutrinos in removing 
energy from the neutron star to a maximum emissivity of
$10^{19}$~ergs/g/s.  
The emission of light saxions 
from the core of the neutron star 
would provide
such an energy-loss mechanism.

Emission rates have been calculated for the emission of other light
particles, 
such as
neutrinos ($NN\rightarrow NN\nu\bar{\nu}$)%
~\cite{Friman:1978zq,Hanhart:2000ae},
axions ($NN \to NNa$)%
~\cite{Hanhart:2000ae,Iwamoto:1984ir,Brinkmann:1988vi,Burrows:1989ah},
scalars ($NN \to NNs$)~\cite{Ishizuka:1990ts},
or Kaluza-Klein (KK) gravitons ($NN\rightarrow NNg$) 
which come about by introducing 
large gravity-only extra dimensions%
~\cite{Cullen:1999hc,Barger:1999jf,Hanhart:2000er,%
Hanhart:2001fx,Fox:2000mt}.
There have also been several calculations for the emissivity of these
particles where it has been assumed that the relevant degrees
of freedom in the core are quarks and gluons, rather than
nucleons~\cite{Iwamoto82,Arndt:2001gz}.

In this work we consider a regime with nucleons as
the relevant degrees of freedom.
In the case of neutrinos, axions, and KK-gravitons 
it has been possible%
~\cite{Hanhart:2000ae,Hanhart:2000er,Hanhart:2001fx} 
by using low energy theorems%
~\cite{Low:1958sn,AdlerDothan}
to relate
the emission rates for soft emission 
to the on-shell nucleon-nucleon scattering amplitudes which can be
extracted from NN scattering phase-shift data.
Such a calculation has the advantage that it is model-independent,
which distinguishes it from the often used one-pion-exchange
approximation where the nucleon-nucleon amplitude is due only to a 
single pion exchange.
A model-independent calculation is possible for the following
reason:
If the energy $\o$ of the radiated particle
is low compared to the momenta of the nuclei 
(low radiation limit) then the main contribution to the emissivity comes 
from bremsstrahlung radiation where the radiated particle couples
to an external nucleon line.
These processes are ${\mathcal O}(1/\o)$, they
exhibit a 
$1/\omega$ pole 
due to an intermediate nucleon being nearly on-shell.
The pole amplitude dominates over
diagrams where the radiated particle couples to an internal line
[which are ${\mathcal O}(\o^0)$].  It is this dominance
that renders the
coupling
of the emitted particle to unknown strong interaction vertices
subleading and enables 
the model-independent calculation.

As explained in Section~\ref{sec:intro}, 
a similar dominance of bremsstrahlung 
is absent for the saxion process $NN\to NNs$.
In each
of the four bremsstrahlung-diagrams there is an infrared pole 
but these cancel in
the sum---diagrams where the saxion radiates off one of the internal
lines are not suppressed and low energy theorems cannot be constructed for
saxion emission.

One way of performing a model-independent calculation would be to use
effective field theory (EFT) methods which have seen some advancement 
in recent years (for a review see, 
for example, \cite{Beane:2000fx}).  
Moreover, it has been shown by 
Beane {\it et al.}~\cite{Beane:2001bc}
that an expansion of nuclear forces
about the chiral limit is formally consistent and 
seems to converge as suggested by numerical evidence.
Beane {\it et al.} 
investigated a toy theory of nucleons interacting with a  
potential consisting of three Yukawa exchanges 
($m_\rho=770$~MeV, $m_\sigma=500$~MeV, and $m_\pi=140$~MeV)
where the masses and couplings where chosen in order to reproduce
the scattering length and effective range in the $^1S_0$ 
nucleon-nucleon
channel.
By treating the pion part of the potential perturbatively 
(the pion coupling $\bar{\alpha}_\pi\propto g_A^2m_\pi^2$)
this model shows convergence up to nucleon momenta
$|\vec{p}|\approx250$~MeV.
While this progress is very encouraging, these methods
are not yet developed far enough to be 
used at momenta
$\gtrsim 300$~MeV as are relevant for supernovae.  
Therefore we will calculate the amplitude for saxion emission
in a $SU(3)$ chiral Lagrangian for baryons
and the pseudo Goldstone bosons associated with
chiral symmetry breaking.
OPE models, based on a $SU(2)$ chiral Lagrangian
have been used in the past for the case of
neutrino,
axion,
and KK-graviton
emission.
Although the OPE captures the relevant physics of the nucleon-nucleon
interaction in the long distance range, it fails to do
so at short distances where the hardcore part of the potential
becomes important.  Although one may naively think
that these contributions
are unimportant for nucleon-nucleon interaction at non-relativistic
energies, this is not true.  Fine-tuning of the underlying parameters
renders the short-distance contributions comparable to the
long-distance ones.

In cases where it is possible (e.g., axions and neutrinos) to calculate
emission rates using both OPE and model-independent techniques,
the two methods
differ by about a factor of three
due to the crude nature of the OPE assertion.  
We expect
our calculation to differ from reality by a similar amount.
This is
sufficient for our purpose and will not change our conclusions
considerably.

Before proceeding we will briefly discuss other saxion emission processes 
and argue why $NN\to NNs$ is dominant.
A possible emission channel is the Compton-like process
$P\g\to Ps$, 
where $P$ and $\g$ refer to a proton and photon, 
respectively.
The leading order $1/\o$ piece in this process cancels between the
two diagrams with uncrossed and crossed external proton lines
and the process contributes at ${\mathcal O}(\o^0)$.
In addition, this process is suppressed by the 
fine-structure constant $1/137$ and the  
proton number fraction $Y_P$ in the neutron star which is
assumed to be small.
Another emission channel,
$N\pi\to Ns$, where $\pi$ is either a thermal pion or
a pion condensate, starts contributing at ${\mathcal O}(\o^0)$
by the same argument.
Also, thermal pions, being bosons, 
obey a Bose-Einstein distribution and their number density is suppressed
by $\exp (-m_\pi/T)\approx 10^{-3}\dots 10^{-2}$ 
for $T\approx 20\dots30$~MeV.
The existence of a pion condensate is
rather uncertain and as such we will ignore its possible effects.  
In short, since the saxion-nucleon coupling is a scalar coupling
it will commute with every other vertex and hence the $1/\o$ pole
will always vanish in the soft-radiation limit.

Before detailing our calculation for $NN\to NNs$,
let us point out several effects
we have left out in this paper.
Throughout the calculation we use in-vacuum amplitudes to calculate
processes in a dense medium.
Whether this approximation is appropriate has been investigated,
for axion emission, in Ref.~\cite{Mayle:1989yx}.
It turns out, that high-density effects are significant and can
modify the emission rate by a factor of about an order of magnitude.

Next,
the emission rate can be suppressed significantly due to
multiple-scattering effects as pointed out by 
Raffelt and Seckel~\cite{Raffelt:1991pw}.
This effect, which is analogous to the 
Landau-Pomeranchuk-Migdal (LPM) effect for electromagnetic
bremsstrahlung by relativistic electrons,
applies if the nucleon collision rate exceeds the
oscillation frequency of the emitted radiation.
If the time to emit radiation is greater than the time between
collisions in the medium then
the nucleons will undergo multiple collisions before 
emission.  
It is the interference between these collisions
that will suppress the radiation.
One incorporates this effect be assigning the nucleon a finite decay
width, $\Gamma$.
This can be done by the following substitution 
in the squared matrix element:
\begin{equation}
  \frac{1}{\o^2}\to\frac{1}{\o^2+\Gamma^2/4}
.\end{equation}
A typical width is given by~\cite{Sedrakian:2000kc}
\begin{equation}
  \Gamma
  =
  aT^2\left[1+\left(\frac{\o}{2\pi T}\right)^2\right]
\end{equation}
where $a\approx 0.2~\text{MeV}^{-1}$.
The LPM effect sets in when $\Gamma\approx\o$, or
$T\gtrsim 10$~MeV.
It suppresses the emission at high temperature, making
the late-time cooling phase of the star more important.

For radiation of particles that couple to the spin of the
nucleon, such as axions or neutrinos, there is an additional
suppression.  
If the correlation length of the nucleons is small
compared to the formation length of the emitted radiation, 
then the emitted
particle couples to an averaged spin since the nucleon-nucleon
interaction can flip spin.  This averaged spin goes to zero
as the nucleon correlation length decreases.
For axion emission this effect reduces the limit on
$f_a$ by about a factor of 2~\cite{Keil:1997ju}.
The saxion, however, is a scalar and does not couple to the
nucleon spin.  We therefore expect no suppression of the emissivity
due to this effect.

\subsection{$SU(3)$ Chiral Lagrangian and Saxion Coupling}
For low-momentum processes chiral-perturbation theory
is the suitable framework to work in.
In contrast to OPE calculations, which are based on $SU(2)$, 
we use in our calculation a $SU(3)$
chiral Lagrangian.  The nucleons in the core of the neutron star
have a typical momentum of $\sim 300$~MeV which suggests the inclusion
of the strange quark in the calculation.
Moreover, as will be explained in more detail below, 
the strength of the saxion-nucleon coupling
depends on the masses
of the octet baryons, which necessitates $SU(3)$.  
As a consequence, 
at tree-level there are not only $\pi^0$ but also $\eta$ exchanges
between nucleons.

The strong dynamics of the octets of pseudo Goldstone bosons
$M$ 
and baryon octet $B$ is described, to lowest order,  
by the Lagrangian 
\begin{eqnarray} \label{eqn:SU3-L}
  {\mathcal L}
  &=& 
  \frac{f^2}{8}\Tr(\partial_\mu \Sigma\,\partial^\mu \Sigma^\dagger)
  +
  \lambda\,\Tr(m_q\Sigma^\dagger+\Sigma m_q^\dagger) 
       \nonumber   \\
  &&+
  \Tr B^\dagger\left(iD_0+\frac{\vec{D}^2}{m_B}\right)B
  +
  D\,\Tr B^\dagger\sigma^i\{A_i,B\}
  +
  F\,\Tr B^\dagger\sigma^i[A_i,B]
\end{eqnarray}
where
$D_\mu=\partial_\mu+V_\mu$
and $m_q$ is the quark mass matrix $m_q=\text{diag}(m_u,m_d,m_s)$.
Furthermore
\begin{eqnarray}
  \Sigma
  &=&
  \exp\left(\frac{2iM}{f}\right),
  \hspace{1cm}
  \Sigma=\xi^2,
  \hspace{1cm}
  M=\frac{1}{\sqrt{2}}\lambda^\alpha\pi^\alpha,  \nonumber \\
  V_\mu
  &=&
  \frac{1}{2}
   \left(\xi^\dagger\partial_\mu\xi+\xi\partial_\mu\xi^\dagger\right),
  \hspace{1cm}
  A_\mu
  =
  \frac{i}{2}
   \left(\xi\partial_\mu\xi^\dagger-\xi^\dagger\partial_\mu\xi\right)
,\end{eqnarray}
and the $\lambda^\alpha$ are the $SU(3)$ generators.  
We use the convention where the
decay constant $f=132$~MeV.

The proton number fraction $Y_p$ is typically small in SN,
especially during
later times in the SN's evolution when most of the protons have
neutronized.
Therefore, and in order to unclutter our formulae, 
we specialize our treatment at this point to $nn$ scattering by
taking $Y_p=0$.
Although we could carry out the calculation for the general case
of a proto-neutron plasma specified by $Y_p$, 
we expect the inclusions of protons to make little difference
in our result. 

To lowest order in the chiral expansion, the contributing diagrams
are those where a single $\pi^0$ or $\eta$ is being exchanged
between two neutron lines.
Therefore,
we need to just keep neutrons $n$ as well as 
$\pi^0$ and $\eta$.
Then the Lagrangian in Eq.~(\ref{eqn:SU3-L}) reduces to
\begin{eqnarray}
  {\mathcal L} \label{eqn:L}
  &=&
  n^\dagger(i\partial_0+\frac{\vec{\nabla}^2}{2M_n})n
  -g_\pi n^\dagger\vec{\sigma}\cdot\vec{\nabla}\pi^0\,n
  -g_{\eta}n^\dagger\vec{\sigma}\cdot\vec{\nabla}\eta\,n
       \nonumber \\
  &&+
    \frac{1}{2}\partial_\mu\pi^0\partial^\mu\pi^0
  -\frac{m_\pi^2}{2}(\pi^0)^2
  +\frac{1}{2}\partial_\mu\eta\partial^\mu\eta
  -\frac{m_\eta^2}{2}\eta^2   
,\end{eqnarray}
where $n$ is the neutron 2-spinor and $M_n$ is the mass of the neutron.
The couplings of neutrons to $\pi^0$ and $\eta$ are 
\begin{equation}
  g_\pi=\frac{D+F}{\sqrt{2}f},\quad
  g_{\eta}=\frac{D-3F}{\sqrt{6}f}
.\end{equation}
Note that the $V_\mu$ coupling has disappeared as it involves a 
baryon-baryon-pion-pion coupling which does not contribute
to saxion emission 
at tree-level.
The meson masses can be calculated in terms of the quark masses,
\begin{equation}
  m_\pi^2=\frac{4\lambda}{f^2}(m_u+m_d),\quad
  m_\eta^2=\frac{4\lambda}{f^2}\frac{m_u+m_d+4m_s}{3}
.\end{equation}

The parameters $D$ and $F$ can be determined from fitting  
to the measured hyperon semileptonic decays. 
This has been done by several authors %
\cite{Jenkins:1991jv,Flores-Mendieta:1998ii,%
      Savage:1997zd,Luty:1993gi,Luty:1993jh}.
At tree level a typical fit gives 
$D=0.79\pm0.10$ and $F=0.47\pm0.07$,
where the errors are highly correlated, 
resulting in the combinations
$D+F=1.26\pm0.08$ and $3F-D=0.65\pm0.21$ \cite{Savage:1997zd}. 
If one includes chiral loop corrections, 
which are ${\mathcal O}(m_s\ln m_s)$,
a typical best fit gives
$D=0.64\pm0.06$ and $F=0.34\pm0.04$ \cite{Savage:1997zd},
where intermediate decuplet baryons have been included.
For our tree level calculation we use typical values of 
$D=0.8$ and $F=0.5$.

To calculate the coupling of saxions to $n$, $\pi^0$ and
$\eta$ from the Lagrangian in Eq.~(\ref{eqn:Ls}) 
we first need to calculate the trace of 
the energy-momentum tensor $T_{\mu\nu}$ from Eq.~(\ref{eqn:SU3-L}).
Making use of the equation
of motion for neutrons,%
\footnote{This is justified since we only calculate diagrams
where the nucleons are external states and therefore on-shell
or nearly on-shell.}
we find
\begin{equation}
  \partial_\mu D^\mu
  =
  T^\mu_\mu
  =
  M_n\bar{n}n
  +2m_\pi^2{\pi^0}^2-(\partial_\mu \pi^0)^2
  +2m_\eta^2{\eta}^2-(\partial_\mu \eta)^2
.\end{equation}
Since the saxion couples to 
$\partial_\mu D^\mu-\sum_{i=1}^3m_i\bar{q_i}q_i$ we need to calculate
the matrix elements of this operator sandwiched between our
low-energy degrees of freedom, neutrons, $\pi^0$, and $\eta$.
Matching
\begin{equation}
  \sum_{i=1}^3 m_i\bar{q_i}q_i
  \longrightarrow
  -\lambda
  \Tr(m_q\Sigma^\dagger+m_q\Sigma)
  =
  \frac{m_\pi^2}{2}(\pi^0)^2+\frac{m_\eta^2}{2}\eta^2
\end{equation}
we get
\begin{equation}
  \langle\pi^0(q')|
    \partial_\mu D^\mu-\sum_{i=1}^3m_i\bar{q_i}q_i
  |\pi^0(q)\rangle
  =
  -2q'\cdot q+3m_\pi^2
\end{equation}
and 
\begin{equation}
  \langle\eta(q')|
    \partial_\mu D^\mu-\sum_{i=1}^3m_i\bar{q_i}q_i
  |\eta(q)\rangle
  =
  -2q'\cdot q+3m_\eta^2
.\end{equation}
Similarly we obtain for the neutrons $n$
\begin{equation}\label{eqn:Mtilde}
  \langle n(p)|
    \partial_\mu D^\mu-\sum_{i=1}^3m_i\bar{q_i}q_i
  |n(q)\rangle
  =
  \tilde{M}
\end{equation}
where $\tilde{M}=687\pm75$~MeV%
~\cite{Nelson:1988sd,Kaplan:2000hh} 
and can be calculated from the
masses of the octet baryons and the
pion-nucleon $\Sigma$-term.
The error for $\tilde{M}$
stems from an estimated 30\% error typical for $SU(3)$ violation. 
Finally, the couplings of the saxion to the nucleon-pion
and nucleon-eta vertices vanish,
\begin{equation}
  \langle n(p')\pi^0(q)|
    \partial_\mu D^\mu-\sum_{i=1}^3m_i\bar{q_i}q_i
  |n(p)\rangle
  =
  \langle n(p')\eta(q)|
    \partial_\mu D^\mu-\sum_{i=1}^3m_i\bar{q_i}q_i
  |n(p)\rangle
  =
  0
.\end{equation}
Putting these interactions into the saxion Lagrangian [Eq.~(\ref{eqn:L-s})] 
we arrive at
\begin{equation}
  {\mathcal L}_s
  =
  \frac{\kappa}{16\pi^2 f_a}
  \left(
    \tilde{M}sn^\dagger n
    +\frac{3}{2}m_\pi^2s(\pi^0)^2-s(\partial_\mu\pi^0)^2
    +\frac{3}{2}m_\eta^2s\eta^2-s(\partial_\mu\eta)^2
  \right)
.\end{equation}

Before calculating the saxion emission amplitude in detail
using the chiral Lagrangian and saxion coupling given above, 
we will do a comparison 
between the cross sections for
saxions and axions in nuclear matter
based on dimensional arguments.

\subsection{Comparison between Saxion and Axion Emissivity}
The coupling of axions to nucleons is 
given by 
\begin{equation}
  \frac{1}{f_a}N^\dagger\vec{\sigma}\cdot\vec{\nabla}a\,N
\end{equation}
so that the lowest order in the $\o$ expansion of the 
cross section for $NN\to NNa$ goes like 
\begin{equation}
  \sigma_a
  \sim
  \frac{1}{f_a^2}
  \frac{\o^2}{\Lambda_\text{QCD}^2}
  \left(\frac{M_n}{\o}\right)^2
.\end{equation}
Here, the $(M_n/\omega)^2$ term is due to
the $1/\o$ pole coming from a nearly on-shell
intermediate nucleon.
Using the Lagrangian for the coupling of the 
saxion to nucleons given in Eq.~(\ref{eqn:Ls})
we can write a similar cross section for $NN\to NNs$:
\begin{equation}
  \sigma_s
  \sim
  \frac{\kappa^2}{(16\pi^2)^2f_a^2}
  \frac{\tilde{M}^2}{\Lambda_\text{QCD}^2}
.\end{equation}
Now, because of the aforementioned $1/\o$ pole cancellation
in the case of the saxion,
the piece of lowest order in $\omega$  
is ${\mathcal O}(\o^0)$.
Comparing the expected saxion and axion emissivities we find
\begin{equation}
  \frac{\dot{\e}_s}{\dot{\e}_a}
  \sim
  \frac{\kappa^2\tilde{M}^2}{256\pi^4M_n^2}
  \simeq
  10^{-2}
.\end{equation}

So on dimensional grounds we expect the saxion emissivity to be 
suppressed be about 2 orders of magnitude
compared to the axion emissivity.

\subsection{Matrix Element}
\begin{figure}[tb]
    \includegraphics[width=\textwidth]{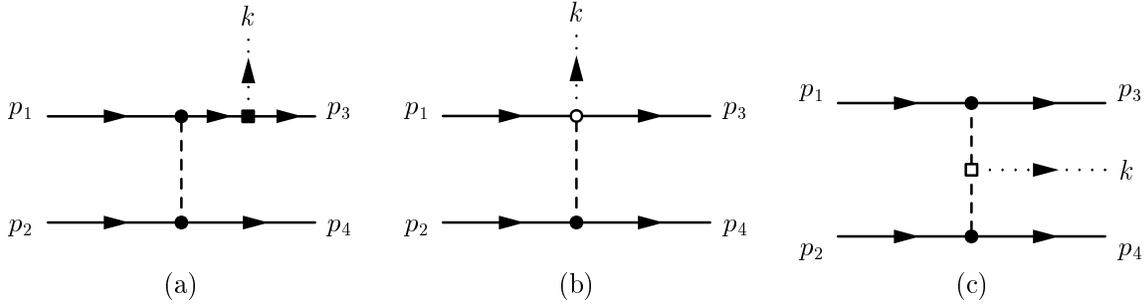}%
    \caption{\label{fig:OPE}
          The three types of diagrams for the process
          $nn\rightarrow nns$.
          Including crossings,
          there are 16 diagrams (8 pion and 8 eta exchanges)
          of the type shown in panel (a),
          4 diagrams of type (c), and no contribution
          from type (b)
          diagrams.
          A solid circle denotes a meson-nucleon-nucleon vertex,
          while a solid (empty) square denotes the coupling of
          the saxion to a nucleon (meson) line. 
          The saxion coupling to the meson-nucleon-nucleon vertex 
          is denoted by an empty circle.
          A dashed line denotes either $\pi^0$ or $\eta$.
          }
\end{figure}

The relevant saxion emission processes are shown in
Fig.~\ref{fig:OPE}.
There is a total of 16 bremsstrahlung-type diagrams 
(4 direct and 4 crossed pion exchanges, 
8 eta exchanges) 
[referred to as type (a) diagrams].
Moreover, there are 4 diagrams where 
the saxion couples to the exchanged meson [type (c)].
The type (b) diagrams do not contribute.

As an example, the diagram pictured in Fig.~\ref{fig:OPE}(a),
where the exchanged meson is the $\pi^0$, gives a contribution of 
\begin{equation} \label{eqn:type-a-example}
  \frac{-i\kappa\tilde{M}g_\pi^2}{16\pi^2f_a}
  (p_4-p_2)_i(p_4-p_2)_j
  \left[\frac{1}{2E_3}\left(\frac{M+E_3}{\o}+1\right)\right]
  \frac{1}{(p_4-p_2)^2-m_\pi^2}
  n^\dagger_3\sigma^in_1\,n^\dagger_4\sigma^jn_2
.\end{equation}
Similarly, the diagram depicted in Fig.~\ref{fig:OPE}(c) for
the $\pi^0$ exchange contributes
\begin{eqnarray}
  &&
  \frac{-i\kappa g_\pi^2}{16\pi^2f_a}
  (p_1-p_3)_i(p_4-p_2)_j
  \frac{1}{(p_1-p_3)^2-m_\pi^2}
  \frac{1}{(p_2-p_4)^2-m_\pi^2}
     \nonumber \\
  && \phantom{asdfasdfasfasfasasdfasdffasf}
  \times
  \left[3m_\pi^2-2(p_1-p_3)(p_4-p_2)\right]
  n^\dagger_3\sigma^in_1\,n^\dagger_4\sigma^jn_2
.\end{eqnarray}

Here, $\sigma_i$ are the Pauli matrices and 
$E_i=M+\vec{p_i}^2/2M$ in the non-relativistic approximation.
Note that in Eq.~(\ref{eqn:type-a-example}) we need to keep
terms
up to ${\cal O}(\o^0)$.  
These will become the main contribution from the type (a) diagrams
as the pole drops out.

The squared amplitude can then be calculated after summing over spins
in terms of the 4-momenta $k=p_2-p_4$ and $l=p_2-p_3$
and is found to be
\begin{eqnarray}
  |{\cal M}|^2 
  &=&
  \sum_{\text{spins}}|{\cal M}_{(a)}+{\cal M}_{(b)}+{\cal M}_{(c)}|^2  
     \nonumber \\
  &=&
  \frac{\kappa^2M^4}{4\pi^4f_a^2}  
  \left\{
    (K_1+K_2)^2k^4+(L_1+L_2)^2l^4 
    -(K_1+K_2)(L_1+L_2)(2[k\cdot l]^2-k^2l^2)
  \right\}
\end{eqnarray}
where 
\begin{equation}
  K_1
  =
  \frac{2\tilde{M}}{E_{CM}}
  \left(
    \frac{g_\pi^2}{k^2-m_\pi^2}
    +
    \frac{g_\eta^2}{k^2-m_\eta^2}
  \right)  
,\end{equation}
\begin{equation}
  K_2
  =
  \frac{g_\pi^2(3m_\pi^2-2k^2)}{(k^2-m_\pi^2)^2}
  +
  \frac{g_\eta^2(3m_\eta^2-2k^2)}{(k^2-m_\eta^2)^2}
\end{equation}
and $L_1$ and $L_2$ are defined analogously 
with $k\rightarrow l$.

In this calculation 
we have assumed the mass of the saxion $m_s$ to be small
compared to the nucleon momenta and
have neglected it.

\subsection{Phase-space integration}
In order to calculate
the emissivity due to saxion $s$ radiation from the two-body
scattering reaction $a+b\rightarrow a+b+s$
in neutron matter, where the labels $a$ and $b$ represent
a neutron,
 we have to integrate over the
phase space of incoming ($\vec{p_1}$, $\vec{p_2}$)
and outgoing ($\vec{p_3}$, $\vec{p_4}$) nucleons
as well as that of the emitted saxion ($\vec{k_s}$).
The formula for the emissivity is
\begin{eqnarray} \label{eqn:emissivity}
  \frac{d{\cal E}_X}{dt}
    &=&
   \int\left[
         \prod_{i=1..4}
         \frac{d^3p_i}{(2\pi)^32E_i}
       \right]
   \int\frac{d^3k_s}{(2\pi)^32\omega}
   f_1 f_2 (1-f_3)(1-f_4)  \nonumber \\
   &&\times (2\pi)^4\delta^4(p_1+p_2-p_3-p_4-k) S\omega|{\mathcal{M}}|^2
\end{eqnarray}
where the functions $f_i$ are Pauli blocking factors given by
$f_i=(\exp[(E_i-\mu)/T]+1)^{-1}$ and
$S=1/4$ is the symmetry factor for $nn$ scattering.

In the soft-radiation limit we can neglect the saxion momentum
$\vec{k_s}$ in the momentum-conserving delta function.
Furthermore, by introducing the total momentum
$\vec{P}=\vec{p_1}+\vec{p_2}=\vec{p_3}+\vec{p_4}$
we can reduce the number of integrations by exploiting spherical
symmetry and momentum conservation to obtain
\begin{eqnarray} \label{eqn:emissivity2}
  \frac{d{\cal E}_X}{dt}
   &=&
   \frac{1}{2^{16}\pi^8}
   \int_0^\infty dP\, |\vec{P}|^2
   \int_0^\infty dk\, |\vec{k}|^2
   \int_0^\infty dl\, |\vec{l}|^2
   \int_{-1}^1d\cos\theta_k 
   \int_{-1}^1d\cos\theta_l
   \int_0^{2\pi} d\phi 
           \nonumber \\
   &&\times
   \frac{\omega^2}{E_1E_2E_3E_4} 
   f_1 f_2 (1-f_3)(1-f_4)
   |{\mathcal{M}}|^2
.\end{eqnarray}
The angles $\theta_k$ and $\theta_l$ are defined by
$\vec{k}\cdot\vec{P}=|\vec{k}||\vec{P}|\cos\theta_k$ and
$\vec{l}\cdot\vec{P}=|\vec{l}||\vec{P}|\cos\theta_l$, 
while $\phi$ is the angle
between the projections of $\vec{k}$ and $\vec{l}$ onto the
$\hat{P}$-plane and is given by
\begin{equation}
  \cos\thCM
  =\cos\phi\sin\theta_k\sin\theta_l+\cos\theta_k\cos\theta_l
\end{equation}
where $\cos\thCM$ is the center of momentum (COM) scattering angle.
Also, $\o$ is constrained by the energy delta function to be
$\o=E_1+E_2-E_3-E_4$.
  
The expression in Eq.~(\ref{eqn:emissivity2})
is valid for all temperatures and chemical
potentials.  In the general regime it must be evaluated
numerically.  

The result simplifies significantly if one assumes either a
degenerate or a highly non-degenerate neutron gas in the star.
Although nuclear matter at densities of a few times nuclear matter density 
and a temperature of $10-50$~MeV is neither degenerate nor
non-degenerate, 
it is nevertheless instructive to calculate 
saxion emissivity in these two limits.

\subsection{Limiting Case: Degenerate Neutron Gas}
For a degenerate neutron gas one can assume
that saxion radiation only arises
from scattering involving neutrons near the Fermi surface in the
initial and final states.
Assuming soft radiation the momentum $|\vec{p_i}|$ of the
neutrons in Eq.~(\ref{eqn:emissivity}) can be set to
$p_F$ causing a decoupling of the energy and angular integrations:
\begin{eqnarray}
    \frac{d{\cal E}_X}{dt}
    &=&
    \frac{p_F^4}{2^{16}\pi^{10}}
    \int_{0}^\infty d\o\,\o^2
    \int dE_1dE_2dE_3dE_4\,
         \d(E_1+E_2-E_3-E_4-\o)f_1f_2(1-f_3)(1-f_4)
                                              \nonumber \\
    && \times
    \int d\O_1d\O_2d\O_3d\O_4
       \d^3(\vec{p_1}+\vec{p_2}-\vec{p_3}-\vec{p_4})
       |{\mathcal{M}}|^2
\end{eqnarray}
where the $\d\O_i$ is the angular integration of the $\vec{p_i}$.

Carrying out the energy~\cite{MoN62,BaP78} and angular integrations 
one finally integrates 
over $\o$ to find
\begin{equation}
  \frac{d{\cal E}_X}{dt}
  =
  \frac{31}{2^{12}945\pi^2}p_FT^6
  \int_0^{\pi} d\a\,\sin\a
  \int_0^{2\pi}d\thCM|{\cal M}|^2
,\end{equation}
where $\cos{\a}=\vec{P}\cdot\vec{p_1}/|\vec{P}||\vec{p_1}|$.
The COM scattering angle $\thCM$ is the angle between the 
projections of $\vec{p_1}$ and $\vec{p_3}$ into the $\vec{P}$ plane.

\subsection{Limiting Case: Non-Degenerate Neutron Gas}
In the limit where the system can be treated as a non-degenerate system the
initial-state distributions become (non-relativistic) 
Maxwell-Boltzmann distributions,
\begin{equation}
  f_i(\vec{p_i})
  =
  \frac{n}{2}
  \left(\frac{2\pi}{MT}\right)^{3/2}
  \exp\left(-\frac{|\vec{p_i}|^2}{2MT}\right)
  \quad\text{for}\quad i=1,2
\end{equation}
where
\begin{equation}
  n
  =
  2\int \frac{d^3p}{(2\pi)^3}\,f_i(\vec{p_i})
\end{equation}
is the neutron number density 
and the factor of 2 is for the two spin states of neutrons.
The final-state blocking factors can be set to unity since basically
the complete final-state phase space is available.
Then, starting from Eq.~(\ref{eqn:emissivity2}), 
the integration over $P$ can be done
analytically, leaving
\begin{eqnarray}
    \frac{d{\cal E}_X}{dt}
    &=&
    \frac{n^2}{2^{14}\pi^{9/2}}
    T^{-3/2}M^{-15/2}
    \int_0^\infty dk
    \int_0^\infty dl
    \int_{-1}^1 d\cos\theta_k
    \int_{-1}^1 d\cos\theta_l
    \int_0^{2\pi} d\phi
                   \nonumber \\
    && \phantom{sffhfhff} \times
    |\vec{k}|^4|\vec{l}|^4\cos^2\theta_{kl}
    \exp
      \left(
        -\frac{|\vec{k}|^2+|\vec{l}|^2+2|\vec{k}||\vec{l}|\cos\theta_{kl}}{4MT}
      \right)    
    |{\mathcal M}|^2
.\end{eqnarray}

\subsection{Numerical Results}
\begin{figure}[tb]
    \includegraphics[width=\textwidth]{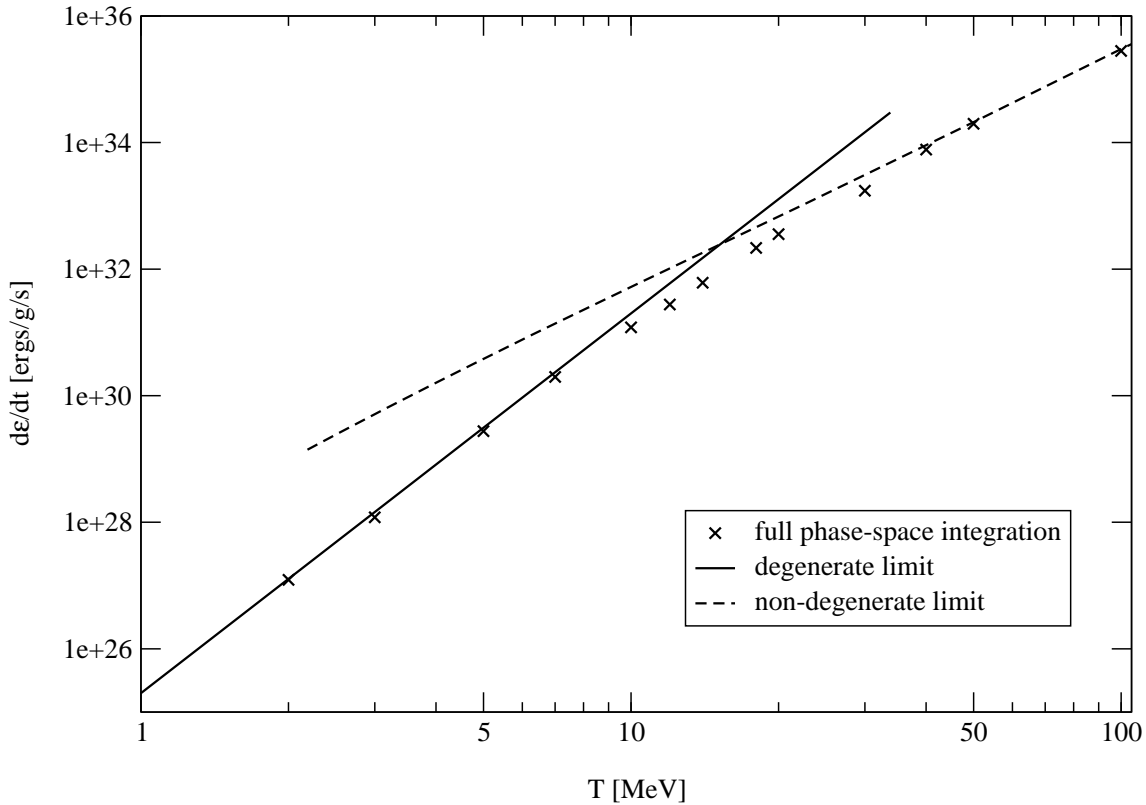}%
    \caption{\label{fig:emissivity}
          Saxion emissivity for fixed $f_a=1$~GeV, $\kappa=1$ 
          at nuclear matter
          density depending on temperature.  
          The crosses are values from the full phase-space integration, 
          the solid and 
          dashed lines denote the result from the degenerate and
          non-degenerate matter calculations.
          }
\end{figure}

We show in 
Fig.~\ref{fig:emissivity}
the emissivity for nuclear density $n=n_0$, where
$n_0=0.17\,\text{fm}^{-3}=1.36\cdot10^6\,\text{MeV}^3$,
and varying temperature 
including the limiting cases of degenerate and non-degenerate matter.
In the figure we have set $\fa=1$~GeV and $\kappa=1$.
As expected, the degenerate approximation is good at low temperature while
the non-degenerate one is better suited to a high temperature regime.

For a typical temperature of $T=20$~MeV and $n=n_0$ the emissivity is
given by
\begin{equation}
  \dot{\e}
  \approx
  \kappa^2
  \left(\frac{\text{GeV}}{f_a}\right)^2
  3.5\times10^{32}
  \frac{\text{erg}}{\text{g}\cdot\text{s}}
\end{equation}
so that, upon using Raffelt's criterion of $10^{19}\,\text{ergs/g/s}$
and Eq.~(\ref{eqn:kappa}),
we can solve for the PQ scale $f_a$
which gives a bound 
of
\begin{equation}
  f_a\gtrsim1.5\times 10^{8}\,\text{GeV}  
\end{equation}
that is close to the bound of $10^9$~GeV coming from 
the axion.
However, if the temperature in the neutron star is as 
high as 50~MeV the
same calculation gives a bound as strong as that 
from axion emission.

The inclusion of the $\eta$, 
typically absent in OPE calculations, 
increases the emissivity by only about 10\% because it is
four times as heavy as the pion. 
Furthermore, we expect the LPM effect to weaken our bound for $T=20$~MeV by
a factor of $\sim 5$.

\section{Conclusions} \label{sec:conclusions}
If both PQ symmetry 
and SUSY are realized in nature, a new scalar particle, 
called the saxion, is introduced 
as a supersymmetric partner of the axion.
The properties of the saxion 
are determined uniquely by SUSY and linked to those of the axion.
Both particles, if they exist and are light, are produced in SN.  
In the past, several calculations using the energy-loss
argument for the axion have been made in order
to place bounds on the PQ scale $\fa$.
In this paper we used saxion emission to bound $f_a$.

The fact that the axion couples to $\Tr F\tilde{F}$ while
the saxion couples to $\Tr F^2$ at an energy scale of $f_a$
causes the two particles to couple very differently 
to nucleons at energies typical to 
neutron stars.  
In the limit that there is a massless quark 
the strong CP problem disappears and 
there is no need for an axion.  
Hence, for a quark mass approaching
zero the axion-nucleon coupling
goes to zero.
In the same limit, however, the saxion does not decouple since it
couples to the QCD field strength.
One might therefore expect that the saxion   
coupling is enhanced over the axion coupling.  
There is also
the added effect of enhancement due to QCD scaling in the
case of the saxion.
On the other hand, saxion emission is suppressed because the
$1/\o$ pole, which is present for the axion, is absent.
Thus it is not clear {\it a priori} how the bounds on
$f_a$ coming from the two particles compare.

Ideally one would like to repeat the model-independent calculation
of the axion emissivity~\cite{Hanhart:2000ae} for the saxion. 
This turns out not to be possible because the $1/\o$ pole
is not present.
Therefore,
low energy theorems, which enable one to relate the emissivity
to the on-shell nucleon-nucleon scattering amplitudes
are no longer applicable and one is forced to
confront the full details of the nucleon-nucleon 
strong interaction.  

In the past the nucleon-nucleon potential has frequently been 
approximated
by an OPE.
This
OPE approximation  
is of a crude nature since it misses 
short-distance physics 
which is
more important than one might naively expect.
Therefore
its
validity is difficult to quantify.
However, 
in the case of neutrino and axion emission
comparisons of the OPE approximation with
a model-independent calculation have shown  
that, surprisingly, the two methods give similar results
to within about a factor of three.
In our calculation of saxion emission
we used an $SU(3)$ chiral Lagrangian, calculating the cross section
to tree-level. 
Since this method is closely related to the OPE approximation, 
one might hope that it
does equally well.

Combining our result for saxion emission with 
the Raffelt criterion we find a bound on $f_a$ that is very
close to the one coming from axion emission.
In view of the significant amount of uncertainty in our calculation
this raises the possibility that saxion emission could
significantly raise the lower bound on $f_a$.
We hope that the exciting prospect of further 
tightening the bound on $f_a$
gives motivation to a model-independent repetition of our
calculation once the
necessary EFT tools become available.

\acknowledgments
We thank Rob Fardon, David Kaplan, Ann Nelson, 
and especially Martin Savage
for interesting discussions.
This work was supported in part by the US Department of Energy under 
Grants No. DE-FG03-97ER4014 (D.A.) and DE-FG03-96ER40956 (P.F.).

\providecommand{\href}[2]{#2}\begingroup\raggedright\endgroup

\end{document}